\documentclass[twocolumn,showpacs,prl,preprintnumbers,amsmath,amssymb,aps,floatfix]{revtex4}
\usepackage{graphicx}
\usepackage{dcolumn}
\usepackage{bm}
\begin{document}
\title{Beyond the Death of Linear Response: $1/f$ optimal information transport.}
\newcommand{\Sss}{\scriptscriptstyle}
\newcommand{\Ss}{\scriptstyle}
\newcommand{\D}{\dysplaystyle}
\newcommand{\T}{\textstyle}
\newcommand{\e}{{\rm e}}
\newcommand{\veps}{\varepsilon}\newcommand{\lgl}{\langle}
 \newcommand{\rgl}{\rangle}
\newcommand{\Vh}[1]{\hat{#1}}
\newcommand{\Aa}{A^1_{\epsilon}}
\newcommand{\Ab}{A^{\epsilon}_L}
\newcommand{\Ae}{A_{\epsilon}}
\newcommand{\finn}[1]{\phi^{\pm}_{#1}}
\newcommand{\ea}{e^{-|\alpha|^2}}
\newcommand{\eb}{\frac{e^{-|\alpha|^2} |\alpha|^{2 n}}{n!}}
\newcommand{\ebbb}{\frac{e^{-3|\alpha|^2} |\alpha|^{2 (l+n+m)}}{l!m!n!}}
\newcommand{\ass}{\alpha}
\newcommand{\as}{\alpha^*}
\newcommand{\fb}{\bar{f}}
\newcommand{\gb}{\bar{g}}
\newcommand{\la}{\lambda}
 \newcommand{\sz}{\hat{s}_{z}}
\newcommand{\sy}{\hat{s}_y}
\newcommand{\sx}{\hat{s}_x}
\newcommand{\sio}{\hat{\sigma}_0}
\newcommand{\six}{\hat{\sigma}_x}
\newcommand{\siz}{\hat{\sigma}_{z}}
\newcommand{\siy}{\hat{\sigma}_y}
\newcommand{\vhsig}{\vec{\hat{\sigma}}}
\newcommand{\hsig}{\hat{\sigma}}
\newcommand{\hH}{\hat{H}}
\newcommand{\hU}{\hat{U}}
\newcommand{\hA}{\hat{A}}
\newcommand{\hB}{\hat{B}}
\newcommand{\hC}{\hat{C}}
\newcommand{\hD}{\hat{D}}
\newcommand{\hV}{\hat{V}}
\newcommand{\hW}{\hat{W}}
\newcommand{\hK}{\hat{K}}
\newcommand{\hX}{\hat{X}}
\newcommand{\hM}{\hat{M}}
\newcommand{\hN}{\hat{N}}
\newcommand{\te}{\theta}
\newcommand{\vze}{\vec{\zeta}}
\newcommand{\vet}{\vec{\eta}}
\newcommand{\vx}{\vec{\xi}}
\newcommand{\vc}{\vec{\chi}}
\newcommand{\hro}{\hat{\rho}}
\newcommand{\vro}{\vec{\rho}}
\newcommand{\hR}{\hat{R}}
\newcommand{\half}{\frac{1}{2}}
\renewcommand{\d}{{\rm d}}
\renewcommand{\top }{ t^{\prime } }
\newcommand{\oz}{{(0)}}
\newcommand{\sint}{{\rm si}}
\newcommand{\cint}{{\rm ci}}
\newcommand{\de}{\delta}
\newcommand{\ep}{\varepsilon}
\newcommand{\De}{\Delta}
\newcommand{\eps}{\varepsilon}
\newcommand{\si}{\hat{\sigma}}
\newcommand{\om}{\omega}
\newcommand{\tr}{{\rm tr}}
\newcommand{\ha}{\hat{a}}
\newcommand{\gam}{\gamma ^{(0)}}
\newcommand{\pe}{\prime}
\newcommand{\BEQ}{\begin{equation}}
\newcommand{\EEQ}{\end{equation}}
\newcommand{\BEA}{\begin{eqnarray}}
\newcommand{\EEA}{\end{eqnarray}}
\newcommand{\sph}{spin-$\frac{1}{2}$ }
\newcommand{\ad}{\hat{a}^{\dagger}}
\newcommand{\add}{\hat{a}}
\newcommand{\spp}{\hat{\sigma}_+}
\newcommand{\smm}{\hat{\sigma}_-}
\newcommand{\fin}[1]{|\phi^{\pm}_{#1}\rangle}
\newcommand{\finp}[1]{|\phi^{+}_{#1}\rangle}
\newcommand{\finm}[1]{|\phi^{-}_{#1}\rangle}
\newcommand{\lfin}[1]{\langle \phi^{\pm}_{#1}|}
\newcommand{\lfinp}[1]{\langle \phi^{+}_{#1}|}
\newcommand{\lfinm}[1]{\langle \phi^{-}_{#1}|}
\newcommand{\lfinn}[1]{\langle\phi^{\pm}_{#1}|}
\newcommand{\z}{\cal{Z}}
\newcommand{\RI}{\hat{{\cal{R}}}_{0}}
\newcommand{\Rt}{\hat{{\cal{R}}}_{\tau}}
\newcommand{\nn}{\nonumber}

\author{Gerardo Aquino$^{1,2}$}\email{g.aquino@imperial.ac.uk}
\author{Mauro Bologna $^{3,4}$ }
\author{Paolo Grigolini$^{3}$}
\author{Bruce J. West$^{5}$}
\affiliation{$^1$Max-Planck Institute for the Physics of Complex Systems,
N\"othnitzer Str. 38, 01187 Dresden, Germany}
\affiliation{$^2$ Division for Molecular Biosciences,  Imperial College London, SW7 2AZ, London, UK}
\affiliation{$^3$Center for Nonlinear Science, University of North Texas, P.O. Box 311427, Denton, TX, 76203, USA}
\affiliation{$^4$ Instituto de Alta Investigaci\'on, Universidad
de Tarapac\'{a}-Casilla 6-D Arica, Chile}
\affiliation{$^5$ Physics  Department, Duke University, Durham, 27708 USA}
\date{\today}
\begin{abstract}
Non-ergodic renewal processes have recently been shown  by several authors to be insensitive to
periodic perturbations, thereby apparently sanctioning the death of linear response, a building block of nonequilibrium statistical physics. We show that it
is possible to go beyond the ``death of linear response" and establish a permanent correlation
between an external stimulus and the  response  of a complex network 
generating non-ergodic renewal processes, 
by taking as stimulus a similar non-ergodic process. 
The ideal condition of $1/f$-noise corresponds to a singularity that is expected to be relevant in several 
 experimental conditions .


\end{abstract}

\maketitle


There has been a surge of interest in understanding the dynamics of complex networks over the past decade with studies ranging from the topology of transportation webs, to the connectivity of communication meshes to the dynamics of neuron networks. Most recently the importance of the nascent theory of  information exchange between complex networks has become evident.

In living neural networks the  connection between function and information transport is studied with experimental techniques of increasing efficiency \cite{physrep} from which an attractive perspective is emerging, {\it i.e.} these complex networks live in a state of phase transition (collective, cooperative behavior), a critical condition that has the effect of optimizing information transmission \cite{beggs}. From the studies of complex networks it is evident that the statistical distributions for network properties are inverse power laws and that the power-law index is a measure of the degree of complexity. Intimate connections exist  between neural organization and information theory, the empirical laws of perception \cite{norwich} and the production of $1/f$ noise \cite{medina}, with the surprising property that  $1/f$ signals are encoded and transmitted by sensory neurons with higher efficiency than white noise signals \cite{lee}. Although   $1/f$ noise production is interpreted by psychologists as a manifestation of human cognition \cite{gilden}, and by neuro-physiologists \cite{yamamoto} as a sign of neural activity, a theory explaining why this form of noise is important for communication purposes does not exist yet.

The well known stochastic resonance phenomenon \cite{stocres} describes the transport of information through a random medium, obeying the prescriptions of  Kubo Linear Response Theory (LRT) \cite{kubo}, being consequently limited \cite{hanggi} to the \emph{stationary equilibrium condition}. 
There are many complex networks that generate $1/f$ noise and  violate this condition:  two relevant examples are blinking quantum dots  
\cite{bqd} and liquid crystals \cite{liquid}.The non-Poisson nature of the renewal processes generated in these examples \cite{brokmann}  is accompanied by ergodicity breakdown and non-stationary behavior  \cite{ergbreak}.

 The response of
   non-stationary 
  networks 
   to harmonic perturbation
  has  recently been found by many authors \cite{barbi, sokolovall, patriarca, karina,weron,sushin} to fade away with time, an effect called \emph{death of linear response} \cite{sokolovall}. This result seems to call into question one of the fundamental theories of statistical physics, that being the fluctuation-dissipation theorem and the resulting LRT of Kubo \cite{kubo}.The intuitive explanation of this effect is as follows. The complex network is prepared and perturbed at time $t= 0$. Experimental preparation  generates a cascade of  events, whose rate $R(t)$ is a decreasing function of time, thereby making the response, which is proportional to $R(t)$,  fade away. Is this a general property, independent  of whether the stimulus is periodic \cite{barbi, sokolovall, patriarca, karina,weron,sushin}?  If  this was a general result, it would be difficult, if not impossible, to explain the communication properties revealed by the recent neuro-physiological literature.

         The purpose of this Letter is to prove that \emph{the death of LRT} actually rests on  an extension of Kubo LRT to the non-stationary condition (NSLRT) and that consequently a non-ergodic system, insensitive to perturbations with a fixed time scale, does respond to perturbations sharing the same non-ergodic behavior.  We shall argue that this important phenomenon explains why  $1/f$ noise is an efficient stimulus for complex systems. 
                           

The  NSLRT rests on the general LRT form \begin{equation}  \label{LRT}
\sigma(t)=\langle \xi _S(t)\rangle =\epsilon \int_{0}^{t}\chi (t,t^{\prime })\xi_P(t^{\prime})dt^{\prime }, 
\end{equation}
where the subscripts $S$ and $P$ denote the ``system'' network and   the perturbing network, respectively. 
Note that $ \left<\xi_S(t)\right>$ is the Gibbs ensemble average over infinitely many  responses $\xi_{S}(t)$ to $\xi_P(t)$ and $\epsilon \ll 1$ is the stimulus strength. We make the simplifying but realistic assumption that  the preparation of S \cite{liquid} does not set a bias on $S$, so that $\left<\xi_{S}(0)\right> = 0$.  The function
 $\chi(t,s)$ is given by \cite{FDT1,FDT2}:
                  \begin{equation}
          \label{FDTb}
        \chi(t,t^{\prime}) = 
 \frac{d\Psi_S(t,t^{\prime})}{dt^{\prime}} = 
R_S(t^{\prime}) \Psi_S(t-t^{\prime}).     \end{equation}
          The function $\Psi_{S}(t,t^{\prime})$ is the autocorrelation function of $\xi_S(t)$, namely, the survival probability of age $t^{\prime}$, and $R_S(t)$ for the case of discrete signals considered here, is the rate 
 at which events are produced by the network $S$ prepared at $t = 0$, \emph{i.e.} the bits per second incoded in $\xi_S(t)$.  This rate is time independent only in the Poisson  case. In the non-Poisson case this rate depends on time, thereby making $\Psi_{S}(t,t^{\prime})$ non-stationary.
 The brand new survival probability $\Psi_S(t) = \Psi_S(t, t^{\prime}$=$0)$, is given by \cite{barbi,FDT1,FDT2}
       \begin{equation}
\Psi_S \left( t\right) = (1+t/T_S)^{1-\mu_S},  \label{nonsur}
\end{equation}
from which the corresponding waiting-times probability density   $\psi_S(t)=-d \Psi_S(t)/dt $ is  derived.
       In the range of parameters $1<\mu_S< 3$  considered here, it is known \cite{brucephysrep} that:
\begin{align}
\label{fadingaway}
      & R_S(t) \approx 
-\frac{\sin{\pi \mu_S}}{T_S} (T_S/t)^{2- \mu_S}
 \;\;\; \;\mbox{ for}\; &1 < \mu_S < 2\\
\label{fadingaway2}
&R_S(t) \approx  
\frac{1}{\tau_S} 
\left[1 + \left(T_S/t \right)^{\mu_S -2}\right]  \;\;\;  \;\mbox{ for}\;\; & 2 < \mu_S <3,
\end{align}
with $\tau_S=T_S/(\mu_S-2)$ the mean value of $\psi_S(t)$. 

 When $\mu_S < 2$ the experimental preparation of $S$ induces a sequence of events,  whose rate $R_S$ tends to vanish for $t\rightarrow \infty$, yielding a perennial out-of-equilibrium condition, and an explanation of the death of linear response \cite{barbi, sokolovall, patriarca, karina,weron,sushin} as well. In  fact, the response to a harmonic perturbation of frequency $f$ is proportional to $1/(ft)^{2-\mu_S}$ \cite{barbi}. 
        In the case $2< \mu_S <3$, on the contrary, the preparation-induced cascade of events, in the limit $t \rightarrow \infty$, becomes stationary and virtually identical to that of a Poisson process. The theoretical analysis of this Letter is done in the asymptotic time regime. Thus, we refer to the case $2< \mu <3$  as \emph{stationary}, in  contrast to the \emph{non-stationary} case $\mu \leq 2$ 
of perennial transition.
    Before proceeding with the use of the NSLRT of Eq. (\ref{LRT}),
we point out some important properties of both  signals $\xi_S(t)$ and $\xi_P(t)$.
If necessary, the signal $\xi_P(t)$ must share the same properties as $\xi_S(t)$ and for simplicity they are both assumed to be
 dichotomous signals with random renewal fluctuations
between the values $+1$ and $-1$. The survival probability in each state is given by Eq. (\ref{nonsur}) 
  with parameters carrying the appropriate index: $T_S, \mu_S$ for $S$ and $T_P, \mu_P$ for $P$.
The spectrum of this type of fluctuating signal, as calculated in Refs. \cite{mb,mirko}, is: 
  \begin{equation} \label{mirkoisverybright}
S(f) \propto L^{\mu-2} f^{\mu-3},
       \end{equation} valid for $\mu <2$, remarkably, even though  a stationary auto-correlation function cannot be defined in this case. In the case $\mu > 2$, $S(f) = A/f^{3-\mu}$, with $A$  independent of $L$, the length of the sequence under study.  
Similarly to the rate of events $R_S(t)$ the  spectral
 intensity per unit time tends to vanish for $\mu < 2$ as an effect of
 increasing $L$. 
The ideal $1/f$
 noise condition, corresponding to $\mu=2$,  generates instead a logarithmic decrease of the spectral intensity with time, and consequently a spectrum
virtually independent of $L$.

Averaging Eq. (\ref{LRT}) over the external fluctuations $\xi_P(t)$ we obtain:
\begin{equation}
\nonumber 
\langle \sigma(t)\rangle =\langle \langle \xi _S(t)\rangle\rangle   = \epsilon
\int_{0}^{t}\chi (t,t^{\prime })\langle \xi _P(t^{\prime
})\rangle dt^{\prime }.  \label{avLRT}
\end{equation}
As previously mentioned,  the NSLRT of Eq. (\ref{LRT}) rests on the preparation of $S$ at time $t=0$. 
 We apply the same preparation condition to $P$, thereby generating the cascades $R_{S}(t)$ and $R_P(t)$ described by Eqs. (\ref{fadingaway}) and (\ref{fadingaway2}), with the appropriate indexing. Under this condition the  relaxation of $\langle \xi_P(t)\rangle$ becomes identical to the survival probability  $\Psi_P\left( t\right) $.
Assuming the condition of Eq. (\ref{FDTb})  we have the following expression for the average response:
\begin{align}\label{fenres}
\langle \sigma(t)\rangle=\eps \int_0^t R_S(t') \Psi_S(t-t')\Psi_P(t')dt' .
\end{align}
The preparation of both $S$ and $P$ makes the average over many realizations of the  response ${\sigma}(t)$ to a given stimulus $P$  vanish for $t\rightarrow \infty$. While we refer the readers to\cite{aquino08} for details, hereby we prove that the intensity of the response of $S$ with $\mu_{S} < 2$ to $\xi_{P}(t)$ does not decay if $\mu_P < 2$.   This is what we mean by going beyond the LRT death, claimed by many researchers \cite{sokolovall,patriarca,karina,weron,sushin}. \begin{table}[h!]
\begin{tabular}{|l||l|l|}
\hline
$\mu_{S\downarrow}$  $\mu_{P\rightarrow}$
 & $\;\; \;\; \;1< \mu_P \leq 2$ & $\;\;\; \;\;\;\;\;\;\; \;\;2<\mu_P< 3$ \\
\hline
 
$1{\textstyle{<}} \mu_S {\textstyle{\leq}}2$  
& $\Phi_{\infty} = \zeta(\mu_S,\mu_P)$$^*$\hspace{0.6 cm}I & $\Phi_{\infty}=0$ \hspace{1.8 cm}II\\
$2{\textstyle{<}}\mu_S{\textstyle{<}}3$ 
 &   $  \Phi_{\infty}=1$ \hspace{1.7 cm} III  & $
\Phi_{\infty}=\frac{\mu_S-2}{\mu_S+\mu_P-4}\;\;\; $ \hspace{0.4 cm}IV \\
\hline
\end{tabular}
\caption{Summary of the asymptotic  values of 
the cross-correlation function $\Phi(t)$. $\;^*$ See Eq. (\ref{ctb}).}
\label{table1}
\end{table}
To prove this important fact we study
the cross-correlation (or I/O correlation)  function
between the system $S$ and the stimulus $P$ 
: $C(t)\equiv \langle \langle \xi_S\left( t\right) \xi _P\left( t\right) \rangle \rangle $  and the mutual information,
which are  used as indicators of aperiodic stochastic resonance \cite{collins}.
Multiplying both sides of Eq. (\ref{LRT}) by $\xi_P(t)$ and averaging over the fluctuations of the perturbation $P$
  we obtain:
\begin{equation}
\Phi(t)\equiv C(t)/\eps
= \int_0^t dt^{\prime }R_S(t')\Psi_S(t-t')\Psi _P\left( t,t^{\prime }\right) .  \label{cross2}
\end{equation}
Note that both  Eq. (\ref{fenres}) and Eq.(\ref{cross2}) depend on the survival probability of network P, but the former depends on the  single time $t^{\prime}$ whereas the latter depends on both $t^{\prime}$ and $t$ and is non-stationary.
We limit ourselves to report the  results for the asymptotic value 
$\Phi_{\infty}$ of $\Phi(t)$.
When  $\xi_{S}(t)$ and $\xi_{P}(t)$ are not stationary, {\it i.e.}  when  $1<\mu_S \leq 2$ and
$1<\mu_P \leq 2$,  Eq. (\ref{cross2}), in the limit
 $t \to \infty$, gives:
\begin{flalign}\label{ctb}
\ & \Phi_{\infty}=\zeta(\mu_S,\mu_P)\equiv
\Gamma(\mu_S+\mu_P-2) \times \\
\nn &   \frac{_3F_2\left[\{\mu_P-1,\mu_P-1,\mu_P+\mu_S-2\},\{\mu_P,\mu_P\},1\right]}{\Gamma(2-\mu_P) \Gamma(\mu_P)^2 \Gamma(\mu_S-1)},
\end{flalign}
where $_3F_2$ is the generalized hypergeometric function.
In the case $2<\mu_P<3$, $\Phi_\infty$ is simply zero.

In the case $2<\mu_S<3$,
inserting into  Eq. (\ref{cross2}) expression (\ref{fadingaway2}) for $R_S(t)$, 
leads to:
\begin{align}
\label{correl_tc}
\Phi(t)  \simeq \int^{t}_0 dt' \frac{1}{\tau_S} \Psi_S(t - t')
\Psi_P (t, t').
\end{align}
Eq. (\ref{correl_tc})  is exact for $t \gg \tau_S$ and for $1<\mu_P \leq 2$ it leads \cite{aquino08} to
$\Phi_{\infty} = 1$, while for $2<\mu_P<3$ it yields:
\begin{align}
\label{statcorr}
\Phi_{\infty}=
(\mu_S-2)/(\mu_P+\mu_S-4).
\end{align}
 \begin{figure}[hbh]
\includegraphics[height=6.8 cm, width=8.8cm]{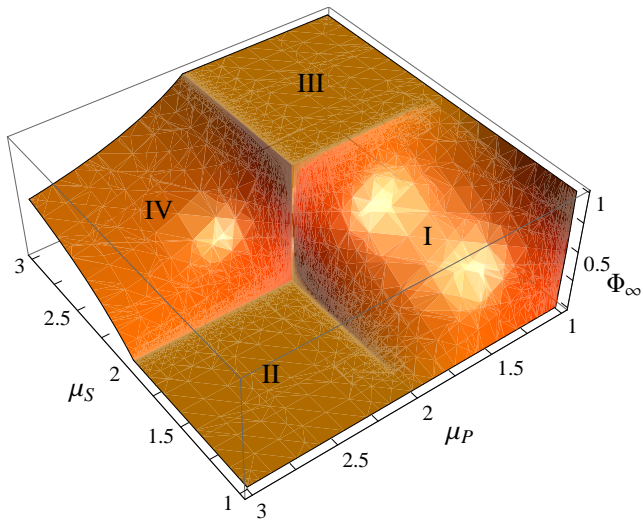}
%
\caption{The asymptotic limit of $\Phi(t)$ is displayed
for  $\mu_S,\mu_P \in ]1,3[$.
The vertex  $\mu_S$=$\mu_P$=$2$  marks the transition  to a condition of maximal input-output cross-correlation.}
\label{fig_1}
\end{figure}
Results are summarized in
 Table \ref{table1}.
For illustrative purposes, we supplement Table I with  Fig. 1,  showing the  3D plot of
the cross-correlation function
 $\Phi_{\infty}$ in the same parameter range:
 Square II and square III correspond to the condition of minimal and maximal correlation, respectively. Intuitively it is so because of the difference of time scales between $S$ and $P$ in such regions. In III fluctuations $\xi_S(t)$ and $\xi_P(t)$ have a finite and an infinite time scale, respectively, thereby allowing
 $\xi_{S}(t)$ to adapt to the stimulus-induced bias so as to yield maximal correlation.  In  II the role of the time scales is inverted,  
the bias induced by $P$  on the longer (diverging) time scale of the process $\xi_S(t)$ is asymptotically averaged out due to the many intervening switching events of $\xi_P(t)$, producing no correlation.  
The vertex $\mu_S=\mu_P=2$, representing a $1/f$-noise system under the stimulus of a $1/f$-noise perturbation,  marks the abrupt transition from  vanishing (square II) to maximal correlation (III). 

Now let us proceed to the demonstration  that the intensity of the response 
$\sigma(t)$ to a single realization of the stimulus
 does not decay, if $\Phi_{\infty} \neq 0$. 
We note that 
 by definition, the non-vanishing $\Phi_{\infty}$ yields:
\BEQ
\nn C(t) \equiv \sum_{i,j} i j \; p\left(\xi_S(t)\textstyle{=}i \big|\xi_P(t) \textstyle{=}j \right)p\left(\xi_P(t) \textstyle{=}j \right) \to \eps \Phi_{\infty},
\EEQ
where the conditional probability for the occurrence of a value of $\xi_S\textstyle{=}i\textstyle{=}\pm 1$, given the occurrence of a  value of $\xi_P\textstyle{=}j\textstyle{=}\pm 1$, has been introduced.
We note that for $t \to \infty$,  on a time scale such that  $\langle \xi_P(t)\rangle \sim\langle \xi_P(0)\rangle t^{1-\mu_P} $
is a second-order quantity, $O(\eps^2)$, we have 
$p\left(\xi_P(t)\textstyle{=}j \right)= 1/2+O(\eps^2)$  and $\Phi(t)=\Phi_{\infty}+O(\eps^2)$.  Thus, due to the symmetry of the considered dichotomous processes:
\begin{flalign}
\label{conditional}
p_t(\xi_S^i|\xi_P^j)\equiv p\left(\xi_S(t)\textstyle{=}i \big|\xi_P(t)\textstyle{=}j \right)&\to \frac{1}{2} + i \, j \eps \frac{\Phi_{\infty}}{2}.
\end{flalign}
In the same long-time scale, Eq. (\ref{conditional}) yields:
\BEQ
\label{finalarg0}
\langle \sigma(t)\rangle_{\pm}
\equiv \sum_{i}p\left(\xi_S(t)\textstyle{=}i|\xi_P(t)=\pm 1\right) i \simeq  \pm \eps
\Phi_{\infty},  
\EEQ 
where the subscript $\pm$ indicates the value of $\xi_P$ at time $t$.
Summing Eq. (\ref{finalarg0}) over the two values of $\xi_P$, gives a total average null response, 
 as expected.
But if the magnitude $|\sigma(t)|$ of the response to a single instance of the input $\xi_P(t)$ is considered instead,  its total average is:
\begin{align}
\label{finalarg}
\langle |\sigma(t)|\rangle=\frac{1}{2} \sum_{\pm}\langle|\sigma(t)|\rangle_{\pm} 
\gtrsim \frac{1}{2} \sum_{\pm}|\langle\sigma(t)\rangle_{\pm}|
\simeq\eps \Phi_{\infty},
\end{align}
where an equality holds if terms of order $O(\eps^2)$ are neglected. 
Thus when $\Phi_{\infty} > 0$, the response $\sigma(t)$ to a single instance of the input $\xi_{P}(t)$ \emph{does not die out and remains proportional to the stimulus}, no matter how large $t$ becomes. Square III is the plateau region of maximal cross-correlation and response, together with the degenerate limit  case $\mu_P=1$ of square I. The term \emph{death of linear response} is appropriate for the vanishing correlation of square II. 
The total average response 
$\langle\sigma(t) \rangle$  always tends to vanish
   for  $t \rightarrow \infty$  for reasons that do not imply a lack of response except in the case of square II. 
\begin{figure}[hth]
\includegraphics[height=4.8 cm, width=7.7 cm]{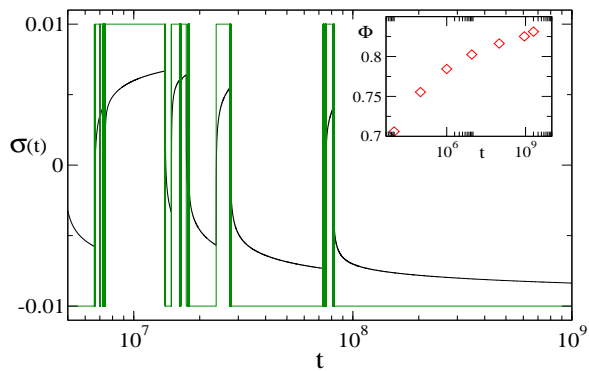}
\caption{Response $\sigma(t)$(black line) to input $\xi_P(t)$ (green square line) rescaled by $\eps$, for $\mu_S$=$1.9,T_S$=$10,\mu_P$=$1.55,T_P$=$9$. {\it Insert}: Average of $\sigma(t) \xi_P(t)$ over $N$=$10^{4}$ inputs, converging to $\Phi_{\infty}$=$0.85$ as predicted by Eq. (\ref{ctb}).(Color online)}
\label{fig_2}
\end{figure}
The ever-lasting response to a single complex stimulus is confirmed numerically  by Fig. 2,  whose insert shows
the correlation emerging from average over many realizations of the stimulus.

The reason for the striking difference between the response to a harmonic perturbation and the response to a non-ergodic stimulus is intimately related to the emergence of $1/f$ noise and to its spectrum described by Eq. (\ref{mirkoisverybright}) which
assigns the  weight $S(f)/L=1/(fL)^{3-\mu_P}$ to the spectral component of frequency $f$ of a non-ergodic stimulus.
As a consequence, the stimulus generates, in time, smaller and smaller frequencies $f$, so as to keep  $1/(fL)^{2-\mu_S}$ ({\it i.e.} the response intensity to frequency $f$ \cite{barbi}) finite, thereby yielding Eq. (\ref{finalarg}). The death of linear response  \cite{sokolovall, patriarca, karina,weron,sushin} is caused by the fact that
stimuli with fixed frequencies cannot cope with the decreasing frequency of the cascade of events of Eq. (\ref{fadingaway}).

We have afforded a compelling proof that the intensity of the single realizations of $\sigma(t)$, with $\mu_S < 2$, does not decay if the perturbation $\xi_{P}(t)$ falls in the same complexity basin ($\mu_P < 2$).
We refer to this phenomen with the term
 {\it complexity management} 
to distinguish it from the term
 {\it complexity matching}
coined in \cite{brucephysrep}
which  implies
 maximum response when $\mu_S=\mu_P$.

Now we argue that $1/f$ stimuli generate the maximum information transport by 
 the mutual information
\BEQ
\label{mutual}
 I(t)=\sum_{ij}p_t(\xi_P^j)p_t(\xi_S^i|\xi_P^j)\log[p_t(\xi_S^i|\xi_P^j)/p_t(\xi_S^i)].\EEQ
From Eq. (12) in fact, it follows that  $I(t \to \infty)\simeq\eps^2 \Phi^2_{\infty}$ and the information transmission rate is obtained by multiplying  $I(t)$
by the input rate \cite{shannon}, given by $R_P(t)$.
If $\mu_{P} > 2$, Fig. 1 shows that $\Phi_{\infty} < 1$.
Although square  III in  Fig. 1  indicates  that  all stimuli with $\mu_{P} \leq  2$ induce maximal correlation,
 $\mu_P<2$ corresponds to
a stimulus   with decaying events rate (input bits/sec) $R_P(t)$.
 So even if a response is produced in this regime,  the rate of  information vanishes in time.
 Only at the crucial condition $\mu_P=2$, of ideal $1/f$-noise , this algebraic decay becomes logarithmic, and, consequently, a steady  and maximal information transmission rate is achieved.

Experimental verification  either on liquid crystals \cite{liquid}  or on ion channels, whose open/close dynamics has been reported to have $1/f$ properties \cite{liebo}, is desirable. In the latter case, using patch-clamp technique, a $1/f$ stimulus can be used as a stimulus and the correlation with the current output analyzed.

In conclusion, the NSLRT proposed in this Letter  \cite{realexp}  explains not only the mystery of the efficient transport of information emerging from the latest theoretical and experimental results in neuro-physiology \cite{beggs,lee}, but many other forms of $1/f$ noise propagations, {\it e.g.} why ecological time series tend to exhibit $1/f$ noise if the underlying abiotic perturbations are $1/f$ noise \cite{halley}.

M. B. and P. G. acknowledge  support from ARO and Welch through grants
W911NU-05-1-0205 and B-1577.

\end{document}